\documentclass[aps,prb,showpacs,preprintnumbers,amsmath,amssymb,superscriptaddress]{revtex4-1}

\usepackage{graphicx}
\usepackage{dcolumn}
\usepackage{bm}
\usepackage{color}
\usepackage{ulem}

\newcommand{\be}{\begin{equation}}
\newcommand{\ee}{\end{equation}}
\newcommand{\bea}{\begin{eqnarray}}
\newcommand{\eea}{\end{eqnarray}}

\newcommand{\expect}[1]{\langle #1 \rangle}

\begin{document}

\title{Selected graphenelike zigzag nanoribbons with chemically functionalized edges. Implications for electronic and magnetic properties}

\author{\bf S. Krompiewski}
\email{stefan@ifmpan.poznan.pl}
\affiliation{Institute of Molecular Physics, Polish Academy of Sciences, 60-179
Pozna\'{n}, Poland}


\date{\today}

\begin{abstract}

It is known that there is a wide class of quasi 2-dimensional graphenelike nanomaterials which in many respects can outperform graphene. So, here in addition to graphene, the attention is directed to stanene (buckled honeycomb structure) and phosphorene (puckered honeycomb structure). It is shown that, depending on the doping, these materials can have magnetically ordered edges. Computed diagrams of magnetic phases illustrate that, on the one hand, \textit{n-type} doped narrow zigzag nanoribbons of graphene and stanene have antiferromagnetically aligned magnetic moments between the edges. On the other hand, however, in case of phosphorene nanoribbons the zigzag edges can have ferromagnetically aligned magnetic moments for the \textit{p-type} doping. The edge magnetism critically influences transport properties of the nanoribbons, and if adequately controlled can make them attractive for spintronics.
\end{abstract}



\maketitle


\section{Introduction}

The great success of graphene has been followed by intensive studies of other two-dimensional nanostructures (2D NS's). Here the main interest is in quasi 2D NS's having either buckled or puckered structures. The well-known representative of the former structure is stanene (\textit{i.e.} 2D Sn), whereas the latter is represented here by phosphorene. There is no doubt that graphene is quite attractive from the point of view of spintronic applications. In particular, its spin diffusion length is quite long, it reveals significant GMR \cite{Hill06,SK09,Yazyev10} and TMR \cite{Seneor12} effects, as well as pronounced non-local spin valve signals, and Hanle spin precessions \cite{Tombros07}. Noteworthy, narrow zigzag graphene nanoribbons have spontaneously magnetized edge atoms as predicted theoretically \cite{Fujita96,Soriano10,SK16} and next confirmed experimentally \cite{Magda14,Li14}.
According to recent studies, the quasi 2D NS's are not inferior to graphene as far as their attractiveness to spintronics is concerned. Band structure and edge magnetism issues in those materials have been recently intensively studied (see Refs.[\onlinecite{Liu11}-\onlinecite{SK18}]). Remarkably, it has been demonstrated experimentally that oxidized phosphorene nanomashes have large magnetic moments at zigzag pore edges \cite{Nakanishi17}.   Following this track, one of the main questions raised in this study is whether a similar effect occurs in the case of narrow zigzag-edge phosphorene  nanoribbons.

\section{Modeling and computational details}

The buckled structure of stanene and the puckered structure of phosphorene are depicted in Figs.1 (b) and (c) (together with the reference graphene structure (a)). Two successive periodicity cells of the infinite zigzag ribbons in the y direction are shown.

\begin{figure}[h!]
\centering \includegraphics [width=0.7\columnwidth, trim=6.9cm 3cm 7.4cm 2.8cm,
clip=true] {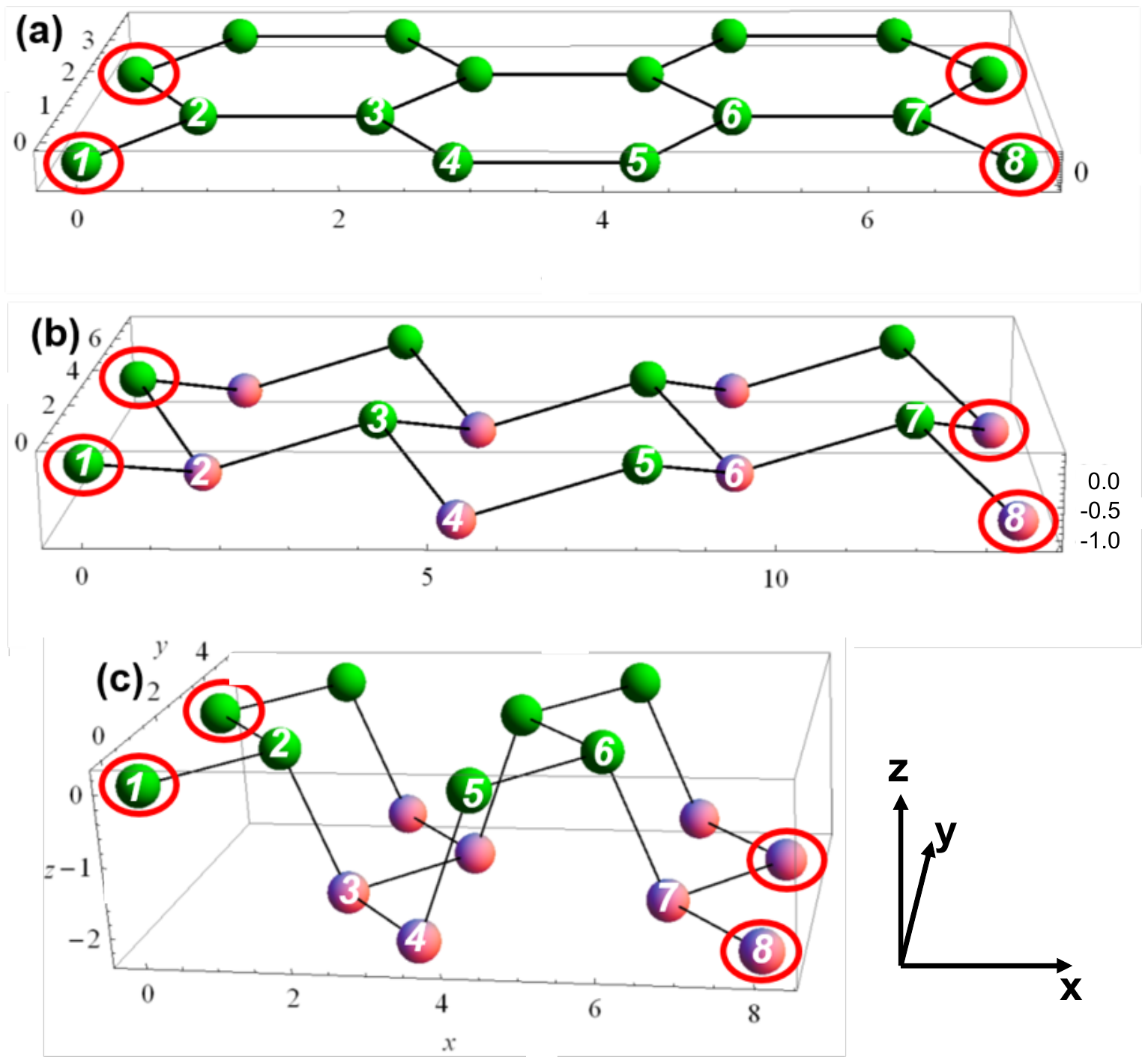}
  \caption{\label{f1}
  The presented nanostructures (NS's) are infinite in the y direction and are two unit-cell (4 zigzag lines) wide. Panels (a), (b) and (c) correspond to graphene (flat NS), stanene (buckled NS) and phosphorene (puckered NS), respectively. Coplanar atoms are marked with the same color and zigzag edge atoms are encircled.}
\end{figure}

The systems are described in terms of the following tight-binding model \cite{SK16,SK18}:

\be
\label{H}
  H = \sum \limits_{ <i j>,\sigma } \,\, t_{ij} \,\, c_{i\sigma}^\dagger c_{j\sigma}+\delta_{0,t_{SO}}H_U^{out}+(1-\delta_{0,t_{SO}})H_U^{in}+H_{dop}+H_{SO},
  \ee

  \be
  \label{U}
  H_U^{out} = -1/2 \sum \limits_{i} \Delta_i \,\, (n_{i \uparrow} - n_{i \downarrow})+1/4 \sum \limits_{i} \Delta_{i} \,\, m_{i}
  \ee

  \begin{eqnarray} \label{Ht}
  H_U^{in} =-U  \sum \limits_{i} [\expect{S_i^+} S_i^- + \expect{ S_i^-} S_i^ + - \expect{S_i^+} \expect{ S_i^-}],
\end{eqnarray}

    \be
    \label{V}
   H_{dop} =  \sum \limits_{i_{edge}, \sigma} \epsilon_{i_{edge},\sigma}  \,\, n_{i_{edge}, \sigma},
  \ee

     \be
    \label{V}
  H_{SO} = i  \; t_{SO} \sum \limits_{ <\!\!<ij>\!\!>} \!\!\!
  \nu_{ij} ( c_{i\uparrow}^\dagger c_{j\uparrow}
  \! -\! c_{i\downarrow}^\dagger c_{j\downarrow}),
   \ee

where the first  term describes interatomic hoppings, $H_U^{out}$ ($H_U^{in}$) describes mean-field Hubbard correlations in the out-of-plane (in-plane) magnetic
configuration. $H_{dop}$ takes into account possible edge functionalization with different chemical groups or by doping. In fact it is a usual on-site potential which is introduced in order to model these effects. For positive (negative) values of $\epsilon$  the occupation number of edge atoms gets reduced (enhanced) depending on whether dopant atoms are more (less) electronegative than the host atoms. Finally, the term $H_{SO}$ describes intrinsic spin-orbit interaction \cite{Kane05}.

The other symbols used above have the following meanings: $n_{i \sigma} = c_{i\sigma}^\dagger c_{i\sigma}$, $S_i^+ =
c_{i \uparrow}^\dagger c_{i\downarrow}$, $S_i^- =
c_{i\downarrow}^\dagger c_{i\uparrow}$, $\Delta_i=U m_i$,
and $\nu_{ij}$ is the Haldane factor  equal to $\pm 1$ depending on wether the path between next nearest neighbor sites (i, j) is (or is not) clockwise.

Local magnetic moments for the out-of-plane and in-plane configurations are given by:

\be
    \label{out}
   m_i \equiv m_i^{out}=<n_{i \uparrow}>-<n_{i \downarrow}>,
    \ee

    \be
    \label{in}
   m_i^{in}=<S_i^+ + S_i^->,
    \ee

    where the expectation values are taken over the ground state of $H$ (see [\onlinecite{SK17}] for details).

On the one hand the out-of-plane magnetic configurations ($t_{SO} \approx 0 $)  will be exemplified with graphene, $t_1=-2.7$ eV, and phosphorene with as many as 5 hopping parameters (in eV) $t_1=-1.22$,  $t_2=3.665$, $t_3=-0.205$, $t_4=-0.105$, $t_5=-0.055$ [\onlinecite{Rudenko14},\onlinecite{SK18}]. On the other hand, in the buckled case, stanene with $t_1=-1.3$eV will be considered, its parameter $t_{SO}=0.0192$eV [\onlinecite{Ezawa15}] is relatively large (Sn is much heavier than C and P) implying the appearance of magnetic anisotropy favoring the in-plane configuration \cite{Lado14,SK16,SK17}. Incidentally, results for the out-of-plane magnetic configuration and the in-plane one do not differ from each other if there is no anisotropy. In what follows, the energy unit is set to $t=\vert t_1 \vert$.

After solving the eigen-problem of the Hamiltonian (\ref{H}), spin-dependent energy bands, numbers of forward propagating modes (ballistic transmission $T^\sigma(E))$ and hence spin-dependent conductance ($G^\sigma$) can be found \cite{SK17}:

\begin{eqnarray} \label{conduct}
G^\sigma &=& \frac{e^2}{h} \int \limits_{-\infty}^{\infty} T^\sigma(E)  (-\partial f(E-\mu)/\partial E) dE.
\end{eqnarray}

\section{Discussion of the results}
\subsection{Magnetic phase diagrams}

It is important to be aware that edge magnetism is strongly energy-dependent (sensitive to the electron filling). Typically it appears at energies close to the chemical potential $\mu=0$, and disappears outside this region. The ground state magnetic configuration can be determined by comparing grand canonical potentials corresponding to different possible magnetic alignments (cf. [\onlinecite{SK17}]).

\begin{figure}[h!]
 \includegraphics [width=0.7\textwidth, trim=4.0cm 8.0cm 4.0cm 8cm,
clip=true] {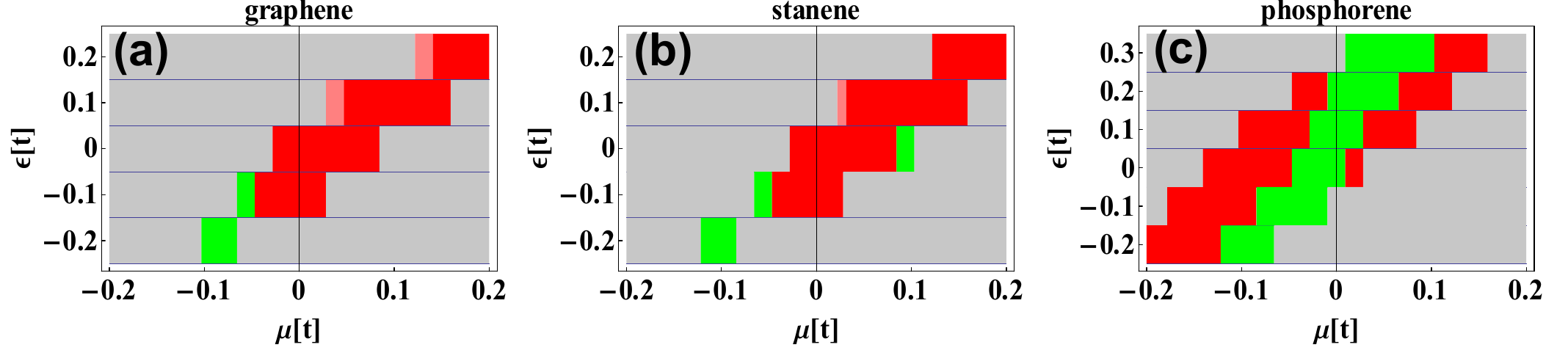}
  \caption{\label{f2}
  The phase diagrams illustrate possible edge magnetization alignments of narrow nanoribbons (5 unit cells  or 10 zigzag lines in width) at room temperature and for various values of the doping parameter $\epsilon$. The colors: gray, red, green and pink denote paramagnetic, antiferromagnetic (\textit{AF}), ferromagnetic (\textit{F}), and 1-edge spin-polarized (\textit{1e}) configurations.}
\end{figure}

Figure \ref{f2} illustrates the phase diagram of possible arrangements of mutual orientations of edge magnetic moments in graphene (panel (a)), stanene (b) and phosphorene (c). Noteworthy close to the chemical potential $\mu=0$ the magnetic phases can appear. In the case of graphene and stanene the antiferromagnetic alignment between the edges is possible for the doping parameter $\epsilon$=0 and -0.1, whereas phosphorene can be ferromagnetically aligned for $\epsilon$=0 and 0.1 and 0.2. It means that the edge magnetism is favored if graphene and stanene are not less electronegative than the dopants, in contrast to phosphorene where it is advantageous if its electronegativity does not exceed that of the dopants.


\subsection{Electrical conductance and edge magnetism of graphene and stanene}

As shown for graphene in [\onlinecite{Magda14}] (experiment and theory), as well as in [\onlinecite{Chen17}] (theory), a typical situation is that narrow zigzag nanoribbons  initially have the \textit{AF} configuration, which gives way to the \textit{F} configuration for increasing width.

As to stanene nanoribbons, the situation resembles that of graphene, as shown in [\onlinecite{Fu17}] the \textit{AF} configuration constitutes the ground state (with edge atom magnetic moments slightly smaller than in graphene). In [\onlinecite{Qi18}] it is also reported that \textit{AF} configuration has the lowest energy, but the energy difference decreases with the increasing width (up to the calculated width of W=16 zigzag lines).

\begin{figure}[h!]
\includegraphics [width=0.7\textwidth, trim=5cm 8cm 2cm 6.5cm,
clip=true] {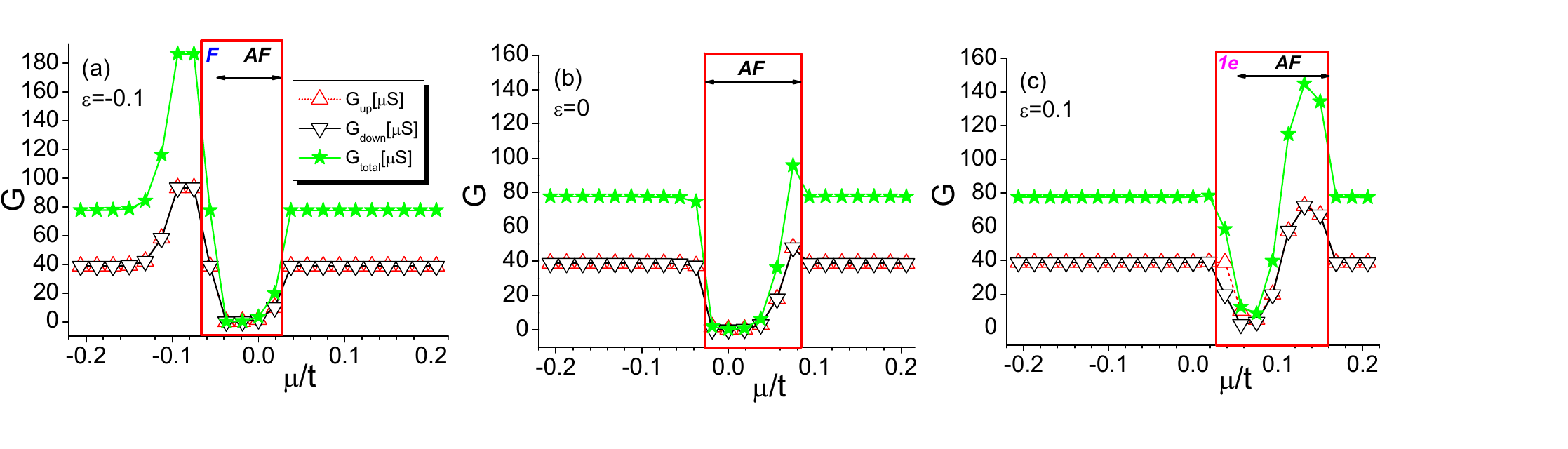}
  \caption{\label{f3}
  Total, up-spin and down-spin conductances (stars, up-triangles and down-triangles) for the graphene NR at T=300K for $\epsilon=$ -0.1, 0 and 0.1 (a, b, and c panels, respectively). Within the red rectangles, the edge magnetism exists.
  }
\end{figure}

\begin{figure}[h!]
\includegraphics [width=0.7\textwidth, trim=4.5cm 7.5cm 2cm 7.5cm,
clip=true] {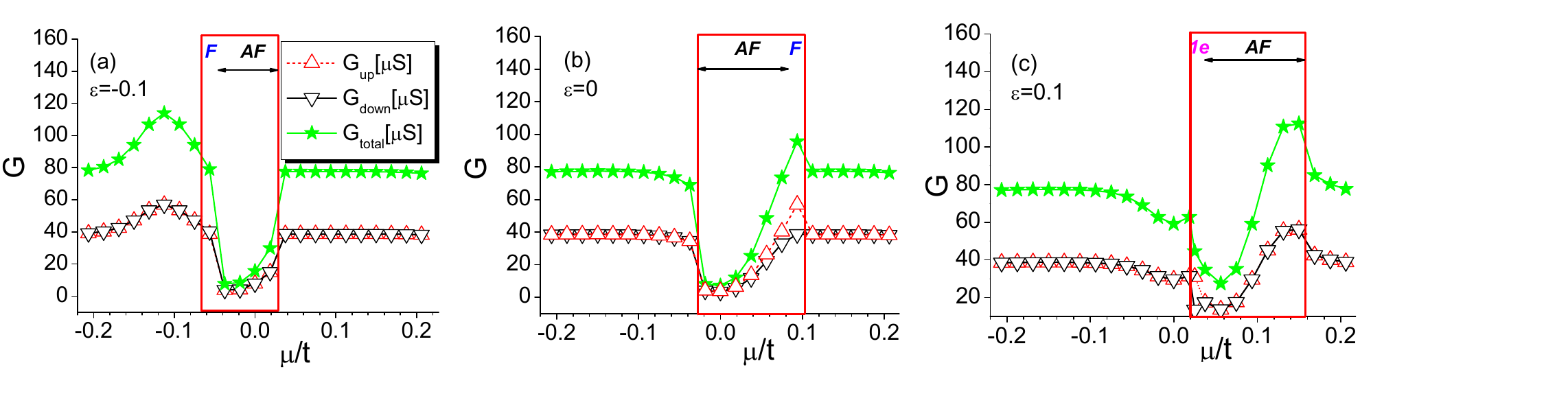}
  \caption{\label{f4}
  As in Fig.\ref{f3} but for stanene NR.
  }
\end{figure}

Figures \ref{f3} and \ref{f4} show spin dependent conductances in the vicinity of $\mu=0$ in graphene and stanene NR's (10 zigzag lines wide). Up- and down-spin conductances are represented by up and down-oriented triangles, and the total conductance - by stars. Rectangles denote the regions where the edge magnetism occurs. In particular, the labels \textit{AF}, \textit{F} stand for antiferromagnetic and ferromagnetic arrangements between the edge magnetic moments, whereas \textit{1e} relates to the situation where essentially only one zigzag edge is magnetic.

Outside the rectangles the edges are paramagnetic and the conductances are relatively high (metallic behavior). In the \textit{AF} region the conductances are initially strongly reduced (semiconductor behavior) but with increasing $\mu$ they can become metallic, whereas in the \textit{F} and \textit{1e} regions some spin splitting of the conductances usually takes place leading to the half-metallic behavior. Interestingly, the \textit{1e} phase (if present) is visible only in a quite narrow energy region on the border between the paramagnetic and magnetic phases.
Depending on initial input parameters the \textit{1e} self-consistent solution leads to the physically equivalent configurations, at the same chemical potential, with the single overdominant peak  located either on the left- or the right-hand side of the ribbon (left-right symmetry is protected). In fact however, in addition to the \textit{1e} spontaneous configuration discussed so far, the appearance of one-edge magnetic configuration can also be due to attaching chemical functional groups just to one edge of a nanoribbon \cite{Zheng08,Ren18}, or keeping one its edge in contact with a magnetic substrate (proximity effect) \cite{Lazic16}. The present approach is appropriate for describing all these cases, in a qualitative way.

\begin{figure} [h!]
  \centering \includegraphics [width=0.7\columnwidth, trim=2.5cm 5.0cm 2cm 5.5cm,
clip=true] {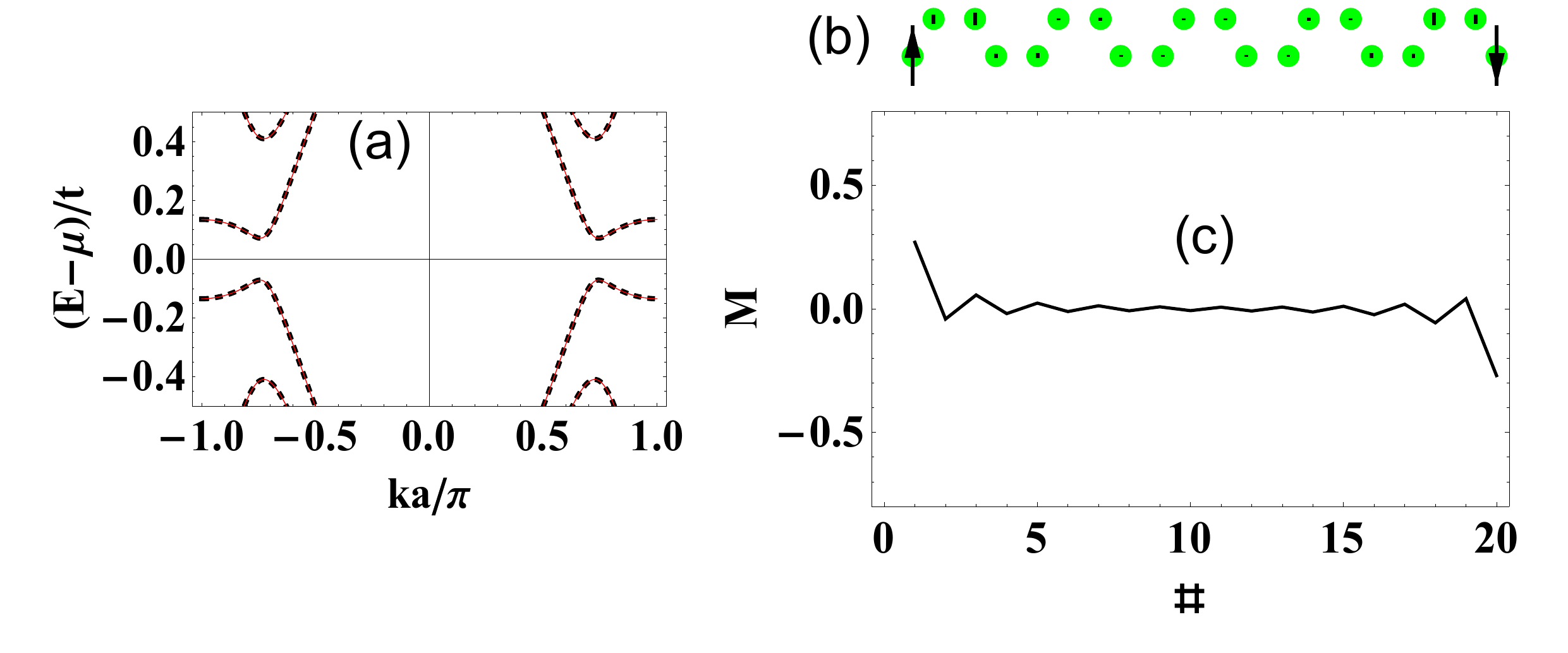}
  \caption{\label{f5}
Graphene with $\epsilon=0$ and $\mu=0$: (a) Low energy up-spin and down-spin energy bands (solid and dashed lies), and magnetic moments in the periodicity unit ((b) and (c)). The atom labels (\#) run over the armchair-type line as in Fig.1.}
\end{figure}

\begin{figure}
 \centering \includegraphics [width=0.7\columnwidth, trim=2.5cm 5.0cm 2cm 5.5cm,
clip=true] {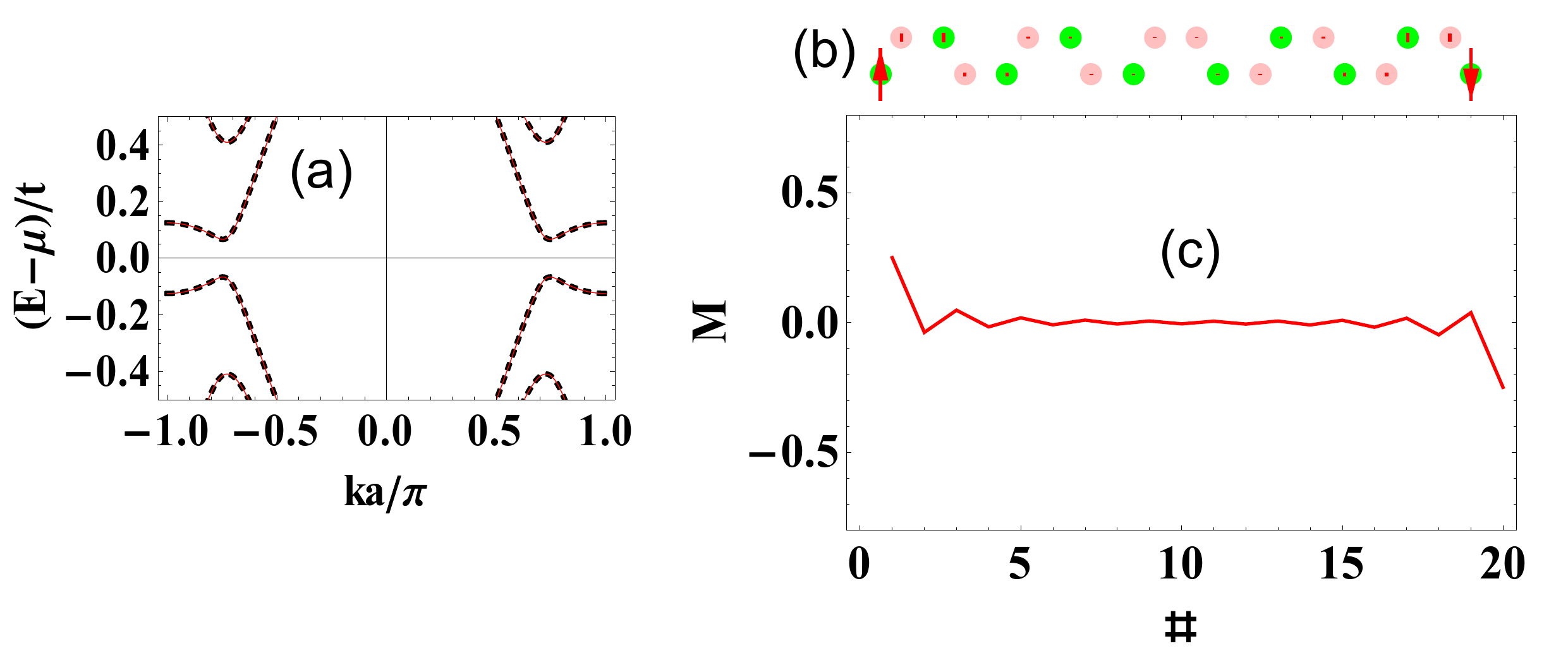}
  \caption{\label{f6}
  As in Fig.\ref{f5} but for stanene ($\epsilon=0$ and $\mu=0$).}
\end{figure}

The key to understanding physical origin of the configurations \textit{AF}, \textit{F} and \textit{1e} is to get a deeper insight into their band structures.
Obviously, low values of the conductances correspond to the situation -  as illustrated in Figs. \ref{f5} and \ref{f6} for the \textit{AF} configuration - with the chemical potential in the energy gap.

\begin{figure}
 \centering \includegraphics [width=0.7\columnwidth, trim=2.5cm 5.5cm 2cm 5.5cm,
clip=true] {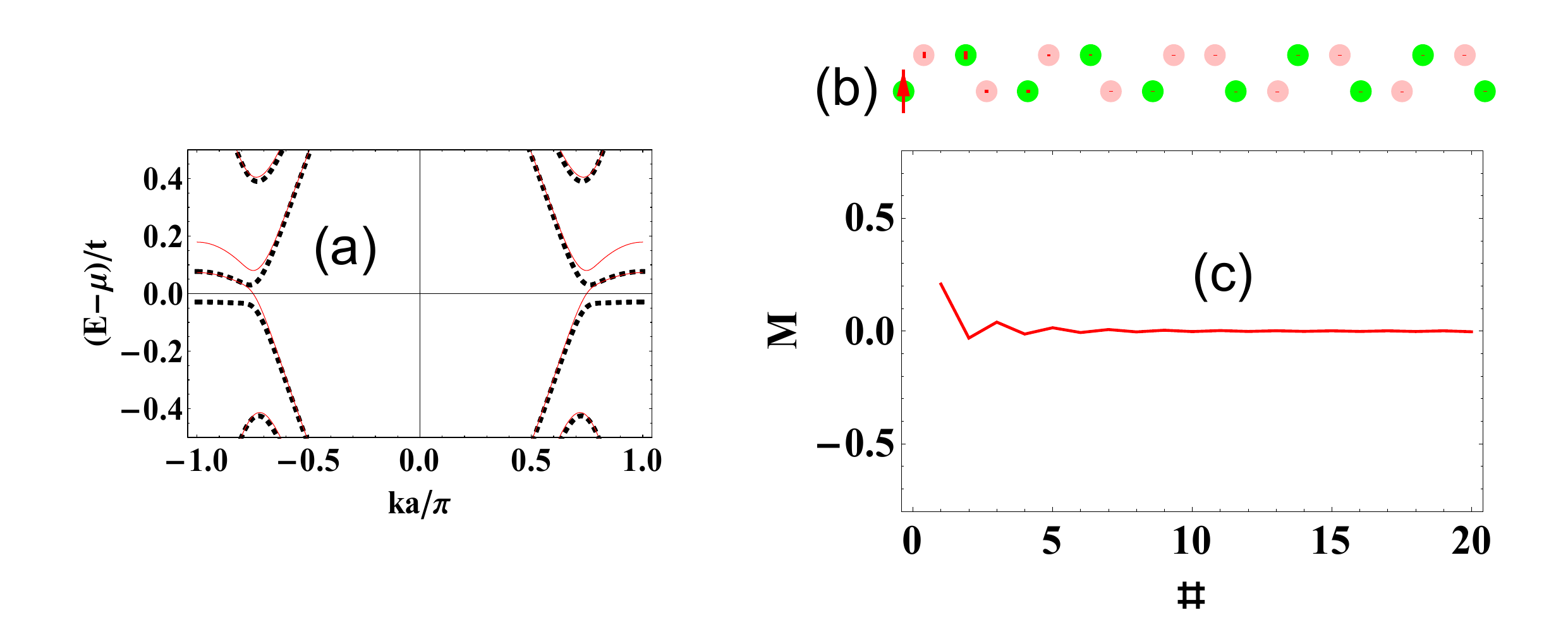}
  \caption{\label{f7}
  The \textit{1e} configuration in stanene for $\epsilon=0.1$ and $\mu=0.025$.}
\end{figure}

Another interesting feature seen in the phase diagram (Fig.\ref{f2}) is the presence of the 1-edge spin-polarized state both in graphene and in stanene. The energy band and the magnetization profile of the latter is shown in Fig. \ref{f7}. These states always have strongly spin polarized conductances and are half-metallic (polarization=100\%).

The magnetic moments in the panels (b) and (c) of Figs. \ref{f5}-\ref{f7} (and
Fig. \ref{f9} below) are presented in the same scale  for comparison purposes.
It is easily seen that in the case of both graphene and stanene the dominating magnetic phase  is the antiferromagnetic one. Moreover, with increasing $\mu$ this region clearly shifts to the right hand side, \textit{i.e.} to higher electron occupancies.

\newpage

\subsection{Electrical conductance and edge magnetism of phosphorene
}

Like graphene, phosphorene can be obtained by mechanical exfoliation from a bulk material and also possesses high carrier mobility. However unlike graphene, phosphorene has got a natural band gap and, for instance, it can be effectively used to construct a field effect transistor \cite{Li-Yu14}. Phosphorene is a very promising material for spintronics, too. In particular, on the one hand, it has been experimentally demonstrated that magnetic moments exist at zigzag-type internal edges of porous phosphorene \cite{Nakanishi17}. On the other hand, there are many theoretical papers reporting on the importance of electronic edge states in phosphorene \cite{Ezawa15,Rahman17,Rahman17,Keshtan15,Yang16}.

It results from the Monte Carlo calculations presented in [\onlinecite{Yang16}] that in the case of phosphorene, narrow zigzag nanoribbons reveal a remarkable edge magnetism and for the width equal to W=6 zigzag lines, the ground state configuration at $\mu=0$ is the \textit{AF} one. Analogous results have been reported in [\onlinecite{SK18}] for W=4 zigzag lines, and it has been shown that for W= 10, the \textit{F} configuration becomes the ground state.

\begin{figure} [h!]
\includegraphics [width=0.7\textwidth, trim=4cm 8.0cm 2cm 6.5cm,clip=true] {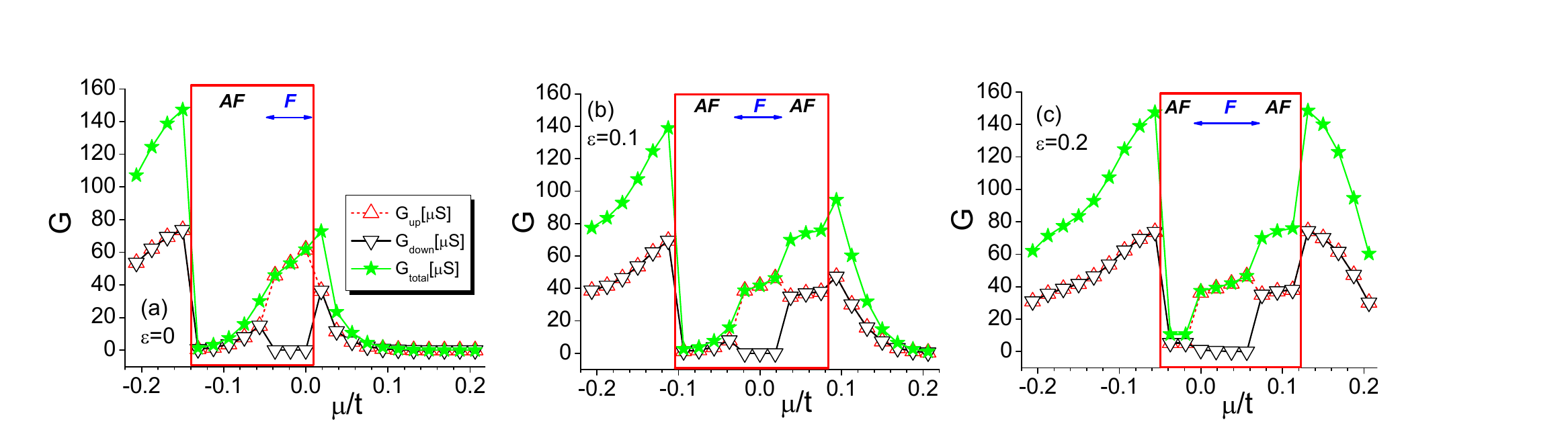}
  \caption{\label{f8}
   Total, up-spin and down-spin conductances (stars, up-triangles and down-triangles) for the phosphorene NR at T=300K for $\epsilon=$ 0, 0.1 and 0.2 (a, b, and c panels, respectively)}
\end{figure}

\begin{figure} [h!]
 \centering \includegraphics [width=0.6\columnwidth, trim=2.5cm 4.7cm 2cm 5.5cm,
clip=true] {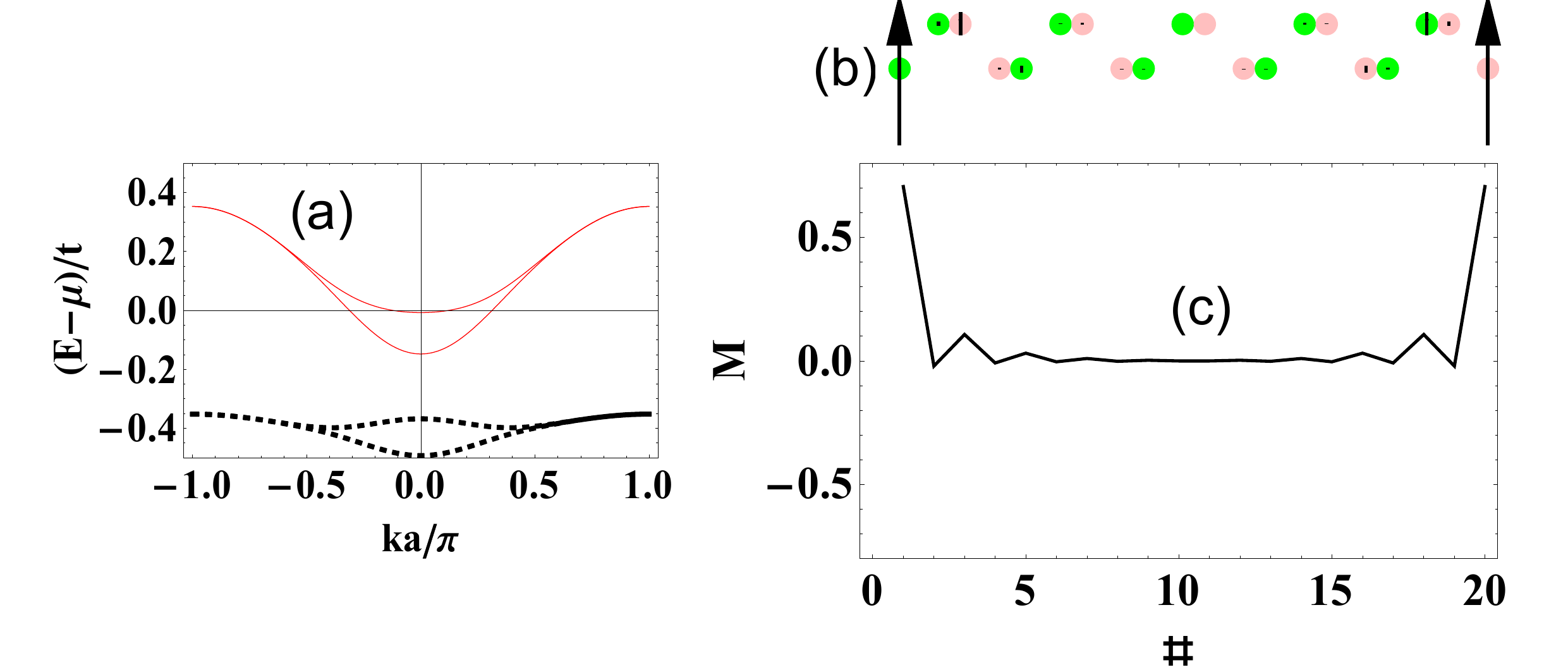}
  \caption{\label{f9}
  Phosphorene with $\epsilon=0$ and $\mu=0$: (a) Low energy up-spin and down-spin energy bands. (b) and (c) magnetic moments in the periodicity unit.}
\end{figure}

Figure \ref{f9} shows that a 10 zigzag lines wide phosphorene nanoribbon has the \textit{F}-type magnetic arrangement for $\epsilon \ge 0$ in the vicinity of $\mu=0$. Moreover in this case the conductance is strongly spin-split and the system is  half-metallic (Fig. \ref{f8}), with $G_{down}=0$ around the chemical potential $\mu=0$. The corresponding energy band structure and the magnetic profiles are presented in Fig.\ref{f9}.

\newpage

The present results agree with the literature data for graphene \cite{Li14,Magda14} and phosphorene \cite{Nakanishi17} where the existence of the edge magnetism was experimentally demonstrated, and it was reported that hydrogenation does not destroy edge magnetism in graphene, but it does in the case of phosphorene (C is more electronegative, whereas P is slightly less electronegative than H). Noteworthy, after the oxidation the situation is reversed and graphene has no longer magnetic edges, whereas phosphorene has magnetic edges (both C and P are less electronegative than O, and p-doping takes place) \cite{Nakanishi17, Boukhvalov11}. Relative electronegativities determine wether n- or p-type doping comes into play, which critically modifies the dangling bonds responsible for the existence of the edge magnetic moments.

It should be also emphasized that the computed zigzag edge magnetic moments in phoshorene are considerably higher than those of graphene and stanene (cf. Fig. 9 with Figs. 5 and 6). This finding - based on the computational method which treats all the nanoribbons under consideration on equal footing - is in qualitative agreement with the experimental results of Refs. [\onlinecite{Li14}] and [\onlinecite{Nakanishi17}], reporting estimates of $M \ge 0.15 \mu_B$ and $M$ up to $1 \mu_B$ for graphene and phosphorene, respectively.
As regards stanene, to the author's knowledge, there are not yet any experiments demonstrating that edge magnetic moments exist in zigzag nanoribbons of this type. It is to be hoped that the present results will inspire work in this direction.


\section{Conclusions}

In summary, by using the self-consistent tight-binding method with Hubbard correlations and spin orbit-coupling, three important atomistic structures of quasi 2D hexagonal zigzag nanoribbons have been studied, namely those with the flat (graphene), buckled (stanene) and puckered (phosphorene) atomic structures. The main focus has been directed to the transformation of electronic and magnetic properties of these materials upon doping or chemical functionalization of the zigzag edges. It  has been shown that all these materials are interesting from the viewpoint of spintronic applications because graphene and stanene can have magnetic edges if n-doped, whereas in the case of phosphorene the edge magnetism can exist for the p-type doping. As shown, the edge magnetism can be effectively controlled by modification of the electron filling factors of the zigzag edge atoms, determining thereby spin-polarized electrical conductances - the essence of spintronics. Phosphorene is the most promising from this point of view because it exhibits half-metallic properties in a relatively wide energy (electron filling) range.


\section{References}


\end{document}